\documentclass[twocolumn]{aastex701}
\usepackage{amsmath,amstext}
\usepackage[T1]{fontenc}
\usepackage{apjfonts} 
\usepackage{multirow}
\usepackage[figure,figure*]{hypcap}
\usepackage{booktabs,caption}
\usepackage{tablefootnote}
\usepackage{hyperref}
\usepackage{subcaption}
\usepackage{mwe}

\begin{document}

\title{Escaping ionizing photons from massive spiral galaxies at $z\sim 1$}

\author[0009-0003-8568-4850]{Soumil Maulick}
\affil{Inter-University Centre for Astronomy and Astrophysics, Ganeshkhind, Post Bag 4, Pune 411007, India}
\email[show]{soumil@iucaa.in}
\author[0000-0002-8768-9298]{Kanak Saha}
\affil{Inter-University Centre for Astronomy and Astrophysics, Ganeshkhind, Post Bag 4, Pune 411007, India}
\email[show]{kanak@iucaa.in}



\begin{abstract}
We report the detection of Lyman continuum (LyC) photons from three massive ($\text{M}_{*}>10^{10}\:\text{M}_{\odot}$) spiral galaxies at a redshift of nearly 1 in the AstroSat UV Deep Field South. Notably, all three systems are viewed at low inclination (i.e., nearly face-on), prompting an investigation into the role of galaxy orientation in the detectability of LyC emission from disk systems. Two of the three galaxies, however, host active galactic nuclei (AGNs), adding complexity to the interpretation of the LyC signal. We present a detailed analysis of the likely star-forming case, and report tentative evidence that a face-on viewing angle may enhance the likelihood of LyC detection in disk galaxies. This represents the first detection of LyC emission from well-characterized spiral galaxies at high redshift, offering a new window into LyC escape mechanisms in such systems. Our findings highlight the need to consider geometric factors and anisotropic escape pathways facilitated by feedback processes alongside more traditional density-bounded scenarios that imply isotropic escape. 
\end{abstract}



\section{Introduction}
The Epoch of Reionization (EoR) marks a major phase transition in the history of the Universe, during which the neutral hydrogen in the intergalactic medium (IGM) was ionized by the first luminous sources. This process is thought to have been primarily driven by low-mass, star-forming galaxies \citep{Robertson22}, though recent evidence suggests that active galactic nuclei (AGNs) may have a non-negligible contribution, particularly during the later stages of reionization \citep{Madau24,Dayal25}.

Despite this progress, the details of how reionization unfolded remain uncertain, most notably because the ionizing photons responsible for it cannot be directly observed due to the high opacity of the IGM \citep{Inoue14} at these redshifts. In contrast, at lower redshifts, where the IGM is more transparent, direct detection of escaping Lyman-continuum (LyC) photons has become possible, leading to the identification of numerous LyC leakers (see for example \citealt{Kerutt23} and references therein). These detections have provided valuable insights into the physical conditions that facilitate LyC escape, though attempts to identify reliable empirical predictions of the LyC escape fraction ($f_{\text{esc,LyC}}$) have significant scatter \citep{Flury22b}.

This complexity has motivated the application of multivariate statistical approaches, such as the analysis by \citet{Jaskot24} to improve predictive power for $f_{\text{esc,LyC}}$. In parallel, high-resolution simulations \citep{Kimm14,Trebitsch17,Ma20} have revealed that LyC escape is a highly dynamic and anisotropic process, regulated by both the spatial and temporal structure of feedback and star formation.

Two theoretical frameworks are commonly invoked to explain the geometry of LyC escape \citep{Zackrisson13}, assuming the source of these photons are young massive stars. In the first, mechanical feedback from stellar winds and supernovae \citep{Heckman11} partially clear channels in the surrounding ISM. However, supernova-driven feedback typically becomes effective only after $\sim$5 Myr, by which point the intrinsic LyC production has already declined significantly. As a result, LyC escape in this case is not only strongly anisotropic, varying by line of sight, as shown in simulations \citep{Cen15,Choustikov23} but also temporally constrained by the evolving interplay between photon production and feedback-driven clearing. {Ex-situ processes such as mergers may also promote the escape of ionizing photons in a manner analogous to feedback-driven ISM clearing, by displacing or disturbing the neutral ISM and creating low-opacity channels. There are indications that mergers can facilitate LyC escape in individual LyC-leaking galaxies at both low and high redshift \citep{LeReste23, Yuan24, RiveraThorsen25}. However, the most prominent LyC leakers in terms of $f_{\text{esc,LyC}}$ remain compact starbursts \citep{Mascia25, LeReste25b}. }
The case of compact starbursts corresponds to the scenario of intense radiation feedback from young, massive stars that fully ionizes the surrounding neutral layers within the Strömgren sphere. This "density-bounded" scenario allows LyC photons to escape more isotropically and is often associated with high ionization conditions, potentially indicated by a large $[\text{O III}] \lambda5007 / [\text{O II}] \lambda3727$ (O32) ratio \citep{Nakajima14} and/or suppressed [S II]$\lambda\lambda$6717, 6731 emission \citep{Wang19} in compact starburst systems.

More recently, two-stage burst models have been proposed to reconcile the temporal offset between the peak LyC production and the onset of effective feedback. In such models, a very young stellar population dominates LyC production, while a near spatially coincident, slightly older population formed in an earlier starburst episode provides the mechanical feedback necessary to clear escape routes via winds and supernovae (see for example \citealt{Enders23,Flury25,Herenz25}).

The escape of ionizing photons from stratified galactic disks, particularly in relation to galaxy orientation has received relatively little attention in recent observational LyC studies. Early efforts to understand this phenomenon were motivated by the need to explain the ionization of the Diffuse Ionized Medium (DIM), also known as the Reynolds layer \citep{Reynolds84}, which lies approximately 1 kpc above the Galactic plane. It was proposed that OB associations could supply the necessary LyC photons to keep this extended layer photoionized \citep{Dove94}. This would then require ionizing photons from young O and B stars to escape their natal birth clouds and propagate over large distances. A key finding from these studies \citep{Dove94} is that star-forming regions with sufficiently high intrinsic ionizing photon production rates in disk systems, tend to become density-bounded in the vertical direction, i.e., perpendicular to the plane of the disk, thereby facilitating the escape of Lyman continuum (LyC) photons along those paths. Additionally, mechanical feedback processes from supernovae and galactic winds may facilitate this process \citep{Dove00}, by preferentially carving dynamic holes or chimneys in the vertical direction. More broadly, this implies that anisotropic LyC escape along favorable lines of sight in extended disk systems may exhibit different integrated observational signatures compared to traditional density-bounded scenarios, such as low O32 ratios \citep{Bassett19} and/or high stellar dust attenuation.

In this work, we report the detection of Lyman-continuum (LyC) photons from three spiral galaxies at redshift $\sim1$ in the Hubble Ultra Deep Field: one appears to be a star-forming galaxy, while the other two host AGNs. Notably, all three systems are viewed nearly face-on. We conduct a detailed analysis of the likely star-forming case, referred to its MUSE HUDF DR2 ID (MUSE ID 16), exploring the scenario in which LyC photons escape along a line of sight that is favorably oriented relative to the galaxy's disk geometry.

\par The paper is organized as follows. In Section~\ref{sec:selection}, we describe our procedure for selecting LyC leaker candidates by combining spectroscopic data from the MUSE Hubble Ultra Deep Field DR2 \citep{Bacon23} with imaging from the AstroSat UV Deep Field South (AUDFs) Survey \citep{Saha24}. Section~\ref{sec:data} details the available archival imaging and spectroscopic datasets for AUDFs\_F15365. These form the basis for our analyses in Section~\ref{sec:methods}, that includes morphological, spectroscopic, and photometric characterization, SED fitting, and estimation of ionizing photon production and escape. In Section~\ref{sec:discussion}, we examine the possibility of LyC photons escaping preferentially along a nearly face-on line of sight, and consider the broader occurrence of face-on LyC leakers. We summarize our findings in Section~\ref{sec:summary}.

All magnitudes quoted are in the AB system \citep{Oke83}. Throughout this work, we adopt the concordance $\Lambda$CDM cosmological model, with $H_0=70\: \text{km}\:\text{s}^{-1}\:\text{Mpc}^{-1}$, $\Omega_m =0.3$, $\Omega_{\Lambda}=0.7$.
\section{Selection Strategy} \label{sec:selection}
We leverage the MUSE Hubble Ultra Deep Field (HUDF) Survey Data Release 2 spectroscopic catalog \citep{Bacon23}, in combination with the AUDFs Survey \citep{Saha24}, to search for candidate Lyman continuum (LyC) leakers in the redshift range $ z\gtrsim0.97$ in the far-ultraviolet (FUV) F154W band of the Ultra-Violet Imaging Telescope (UVIT). This lower redshift limit is set by the requirement that the red end of the F154W bandpass falls within the bluer side of the Lyman limit for these sources. We also adopt an upper redshift limit of $z \sim 1.4$, that ensures the [O II]$\lambda\lambda$3727, 3729 doublet, typically the strongest emission feature in the observed optical regime at these redshifts, remains well within the MUSE spectral coverage. Using the Advanced MUSE Data Products database (AMUSED\footnote{\href{https://amused.univ-lyon1.fr/}{https://amused.univ-lyon1.fr/}}), we identify an initial sample of 286 galaxies that satisfy these redshift criteria.
We visually inspect all 286 galaxies using a combination of high-resolution JWST NIRCam imaging from the JWST Advanced Deep Extragalactic Survey (JADES, \citealt{Eisenstein23,Rieke23}) and the UVIT F154W imaging (including the segmentation map) from the AUDFs survey \citep{Saha24}, and classify each source using a flagging system based on proximity and the UVIT F154W detection:
\begin{itemize}
    \item \textbf{Flag 0}: Objects that have neighboring sources within a 1.2$''$ radius, comparable to the resolution of the UVIT F154W band, in the NIRCam images, and for which there is "some" level of detection in the F154W filter within this radius. Here, "some" refers to cases where the centroid of the UVIT-detected object may lie outside the 1.2$''$ radius, but its spatial extent, as defined by segmented pixels in the segmentation map constructed in \citet{Saha24}, partially overlaps with the MUSE position. These cases present ambiguity due to possible foreground contamination in LyC and are hence excluded from further analysis. This is the largest class, comprising 189 objects.
    \item \textbf{Flag 1}: Objects for which there is no UVIT F154W detection, including even partial overlaps, within a 1.2$''$ radius of the object's center as seen in the NIRCam image. These are considered clean non-detections, and constitute our LyC non-detection sample, comprising 92 galaxies. We examine the morphological and orientation properties of this sample in Section \ref{sec:inclination}.
    \item \textbf{Flag 2}: Objects with no nearby neighbors within 1.2$''$ in NIRCam and with a clear detection in the UVIT F154W band. These represent our candidate LyC leakers, and total 5 galaxies. Out of these five LyC leaker candidates, two were previously reported in our earlier works \citep{Maulick24,Maulick25}, which was based on the HST grism CLEAR survey \citep{Simons23}. These are MUSE ID 899 (AUDFs\_F13753, \citealt{Maulick24}) and MUSE ID 13 (AUDFs\_F16775b, \citealt{Maulick25}). The three new candidates identified in this work are MUSE ID 8, 16, and 893 in the MUSE HUDF DR2 catalog.  
\end{itemize}
So far, we have not imposed any morphological constraints in our selection. It is therefore serendipitous that the three new candidate LyC leakers we identify in this work exhibit spiral-like morphologies and appear to be viewed nearly face-on (see Figures~\ref{fig:stamp_images} and \ref{fig:MUSEID8_893}). All three candidates are detected in the UVIT near-ultraviolet N242W band, which probes the non-ionizing UV continuum at these redshifts. For MUSE ID 16 and MUSE ID 893, we identify unique counterparts in the UVIT F154W catalog of \cite{Saha24}, with the IDs AUDFs\_F15365 and AUDFs\_F14141, respectively. Their signal-to-noise (S/N) ratios in the F154W band, computed with an aperture of radius 1.6$''$, taken from the catalog presented in \citet{Saha24}, are 3.17 and 3.25, respectively. MUSE ID 8, however, is not formally detected in the F154W catalog. Nonetheless, forced photometry performed at the optical centroid using a 1.6$''$ radius aperture yields a signal-to-noise ratio exceeding 3 ($\sim3.6$). Based on both this S/N and our visual inspection, we consider MUSE ID 8 a valid detection in the F154W band.

Among these three, two galaxies—MUSE ID 8 and MUSE ID 893 are detected in X-rays in the 7 Ms Chandra Deep Field-South survey \citep{Luo17}. Based on the classifications of the \cite{Luo17} catalog, MUSE ID 893 is identified as an AGN, while MUSE ID 8 is labeled as a galaxy. However, both sources exhibit broad H$\alpha$ emission features in their HST G141 grism spectra, as observed in the CLEAR survey \citep{Simons23}, which further supports the presence of AGN activity. In contrast, MUSE ID 16 which is the primary focus of this work is not detected in X-rays and does not show any broad H$\alpha$ feature in its grism spectrum, effectively ruling it out as a Type 1 AGN. Due to limited spectral coverage, we are unable to perform a full Baldwin, Philips, \& Terlevich (BPT, \citealt{Baldwin81}) diagnostic for these sources.

Given the likely AGN nature of MUSE ID 8 and MUSE ID 893, we focus the rest of this work on MUSE ID 16, 
which exhibits no strong AGN signatures and is therefore better suited for exploring LyC escape from star-forming regions in a disk geometry, the primary focus of this study. The two AGN-associated LyC candidates are thus shown in Figure~\ref{fig:MUSEID8_893} and discussed further in Appendix~\ref{sec:AGN_leakers}.
While this work centers on a likely non-AGN source, we note that geometric considerations relevant to LyC escape from disk-like structures may still apply in AGN-hosting systems. More recently, empirical evidence from LyC-leaking AGNs at redshifts $2.3\lesssim z\lesssim3.7$ \citep{Smith24} has shown that LyC escape from AGNs is a complex process, challenging the common assumption of a near-100\% escape fraction. We leave a detailed exploration of this to future work. 

We return to the larger sample that also includes the non-detections in Section \ref{sec:inclination} to investigate the potential role of galaxy orientation with respect to the observer, in the detection of LyC.

\section{Imaging and Spectroscopic Data} \label{sec:data}
We utilize far- and near-ultraviolet imaging of the GOODS-South field from the AUDF South survey \citep[see][]{Saha24}, obtained with the UVIT. Specifically, the F154W and N242W filters probe the rest-frame Lyman continuum (LyC) and non-ionizing FUV emission, respectively, from the target object. In addition, we incorporate high-resolution imaging from the Hubble Space Telescope (HST), using data from WFC3/UVIS, ACS/WFC, and WFC3/IR as part of the Hubble Legacy Field (HLF)\footnote{\href{https://archive.stsci.edu/prepds/hlf/}{https://archive.stsci.edu/prepds/hlf/}} program \citep{Illingworth16,Whitaker19}. We also include JWST NIRCam imaging from the JADES survey \citep{Eisenstein23,Rieke23} \footnote{\href{https://archive.stsci.edu/hlsp/jades}{https://archive.stsci.edu/hlsp/jades}} and adopt JWST MIRI photometric measurements from the MIRI Deep Imaging survey (MIDIS, \citealt{Ostlin25}). Together, the HST and JWST imaging span a rest-frame wavelength range of approximately 0.1–2.7~$\mu$m. Finally, in Section \ref{sec:morph}, we also analyze archival HST ACS/Solar Blind Channel (SBC) imaging in the F150LP filter, retrieved from MAST under program ID 10403 (PI: Teplitz).
\par MUSE ID 16 lies in the 141-hour MUSE eXtremely Deep Field (MXDF; \citealt{Bacon23}). We examine the MUSE spectrum of the galaxy in Section \ref{sec:spec}. The H$\alpha$ and Pa$\alpha$ emission from this galaxy is captured by slitless spectroscopic data from the 3D-HST program (HST G141 grism, \citealt{Momvheva16}) and the FRESCO (JWST NIRCam F444W grism \citealt{Oesch23}) surveys (Section \ref{sec:spec}). 

\begin{table}
\centering
\nolinenumbers
\caption{Properties of MUSE ID 16. All reported line fluxes are the observed values and are in units of $10^{-17}\text{erg}\ \text{s}^{-1}\ \text{cm}^{-2}$. The final observed H$\alpha$ line flux however is obtained after correcting for [N II] contamination (Section \ref{sec:spec}). The escape fraction estimates have been corrected for IGM absorption.}

\begin{tabular}{p{3cm}p{3cm}}
\hline
Column name & Value \\ \hline

MUSE HUDF DR2 ID                                                              &      16                                                                                                                          \\

UVIT F154W ID                                                              &      AUDFs\_F15365                                                                                                                                                                                                   \\

R.A., decl. (J2000)                                                                   & 53.1659, -27.7816                                                                                                            \\

$z$                                                                   & 1.097                                                                                                                                                                              \\

F154W (LyC) S/N &          3.17 \\
F154W mag (aperture corrected) &  $26.52\pm0.34$ \\

${M}_{\text{UV}}$ (at 1500 \AA) & ${-18.34 \pm 0.05}$ 
\\
$\text{H}\alpha$ line flux                                                                  & {$8.07\pm 0.55$}                                                                                                               \\
Pa$\alpha$ line flux                                                               & {$1.80\pm0.3$} 
 \\

{[O II] $\lambda3727$} line flux                                                           & {$1.00\pm0.01$}                                                                                            \\

[Ne III] $\lambda$3870 line flux                                                               & {$0.06\pm0.01$} 
 \\

{$\text{log}_{10}(\text{M}_{*}/\text{M}_{\odot})$}                                                           & {10.33$\pm$0.03}                                                                                  \\
{$\text{SFR}_{\text{SED,10\:Myr}}$ ($\text{M}_{\odot}\ \text{yr}^{-1}$)}                                                          & {21.13$\pm$ 6.11} \\

$\text{SFR}_{\text{Pa}\alpha}$ ($\text{M}_{\odot}\ \text{yr}^{-1}$)
             & {$9.72\pm 1.78$}\\
$\text{Age}_{\text{main,SED}}$ (Gyr) & $1.4\pm0.6$ \\
$\text{Age}_{\text{burst,SED}}$ (Myr) & $16.2\pm4.5$ \\
$\beta_{\text{obs}}$ & $-0.31 \pm 0.07$ 
\\ 
{${E(B-V)}_{\text{neb,H}\alpha/\text{Pa}\alpha}$}                                                               & {$0.34\pm 0.20$}\\
{${E(B-V)}_{\text{neb,SED}}$}                                                               & {0.30}\\
{$\text{log}_{10}(\xi_{\rm{ion}}/\text{Hz\: erg}^{-1})$ }                                                            & {${25.14\pm 0.06}$} \\                    
$f^{\rm{H\alpha}}_{\rm{esc}}$                                                            & {${0.39\pm0.15}$} \\                                                        
$f_{\rm{esc,SED}}$                                                      & {$0.28\pm0.13$}                                                                                        \\

\label{tab:MUSE16_prop}

\end{tabular}
\end{table}

\section{Methods and Analysis} \label{sec:methods}

\subsection{Morphological analysis} \label{sec:morph}
\begin{figure*}[ht!]
\nolinenumbers
\includegraphics[width=1\textwidth]{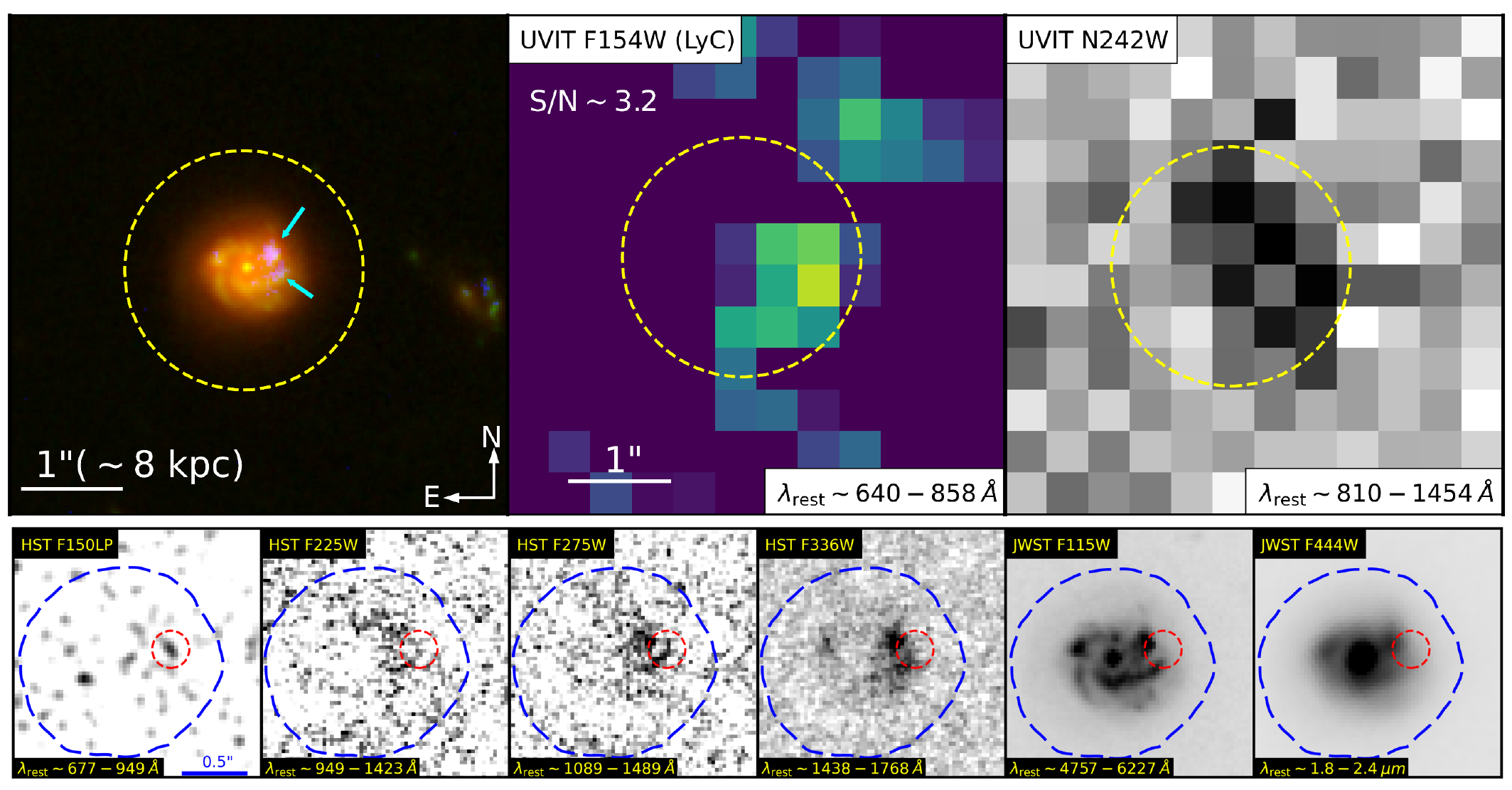} 
\caption{Multiband imaging of 
MUSE ID 16. The RGB image ($5''\times5''$ size) is composed using the HST F336W (blue), JWST F115W (green), and JWST F444W (red) band images, which are also shown individually in the bottom panel in a zoomed-in view ($2''\times2''$ size). The yellow aperture in the upper panel is centered on the F444W band image centroid and is of radius 1.2$^{''}$. The cyan arrows in the RGB image highlight the blue star-forming regions identified within the galaxy.
The first three panels from the left in the bottom row show 
MUSE ID 16 in the bluest available HST bands, covering a rest-frame wavelength range of 677–1489 \text{\AA}, while the last two panels display the galaxy in JWST/NIRCam F115W and F444W bands, probing optical to near-infrared rest-frame wavelengths. The red aperture of radius 0.15$''$ marks the emission complex observed in the HST F150LP band, which is likely associated with the galaxy. The signal within this aperture is detected at a significance of $\sim2.8\sigma$ (see Section \ref{sec:morph} for details). The blue dashed contour marks the 15$\sigma$ isophote in the F444W image and is shown for visual reference.
Note that the UVIT F154W and the HST F150LP band images displayed have been smoothed using a Gaussian kernel of FWHM 3 pixels for visual clarity.} 
\label{fig:stamp_images}
\end{figure*}

In Figure \ref{fig:stamp_images} we display 
MUSE ID 16 in some of the HST and JWST NIRCam bands. The morphology appears markedly different in the HST bands compared to the JWST bands (bottom panel of Figure \ref{fig:stamp_images}): the HST WFC3/UVIS bands (F225W, F275W and F336W), which trace rest-frame non-ionizing FUV emission, reveal clumpy star-forming regions while in the {JWST NIRCam F115W band that probes the rest-frame optical emission of the galaxy}, it exhibits a well-defined spiral structure. The clumps seen in the HST UV bands correspond to regions of recent star formation (indicated by the cyan arrows in the upper panel of Figure \ref{fig:stamp_images}), in contrast to the more dust-enshrouded central core. The stellar disk of the galaxy extends well beyond the clumpy star-forming regions, as evidenced by the 15$\sigma$ F444W boundary (blue dashed contour in the bottom panel of Figure~\ref{fig:stamp_images}) that traces the rest-frame $\sim2.1 \:\mu$m light, suggesting the presence of an underlying older stellar population.
In the second panel in the top row, we display the LyC signal from this object detected at a $\sim3.2\sigma$ significance above the background fluctuations in the UVIT F154W band. While the centroids of both UVIT band images (F154W and N242W) appear offset toward the clumpy star-forming regions, we find that after convolving both the HST F275W and JWST F444W images with the UVIT F154W PSF and resampling them to the UVIT pixel scale, the distinct regions visible in the high-resolution data become indistinguishable. 
\par The HST ACS/SBC F150LP filter probes approximately the same rest-frame wavelength range as the UVIT F154W filter. However, due to the extended red cutoff of the F150LP bandpass, it also admits a contribution from non-ionizing photons.  
In the HST F150LP image (first panel in the bottom row of Figure \ref{fig:stamp_images}), we identify emission roughly coincident with that seen in the bluest HST UVIS filters. We mark this emission complex using the red aperture in Figure \ref{fig:stamp_images}. To assess its significance, we measure the flux within the red aperture (radius=0.15$''$) and employ a Monte Carlo method following \citet{Maulick24}. Specifically, we place non-overlapping random apertures of the same size across a 10$''$ $\times$ 10$''$ region centered on the source. By analyzing the resulting flux distribution and fitting a Gaussian, we determine that the flux within the red aperture corresponds to a $\sim2.8\sigma$ detection. Interestingly, the emission peak in the F150LP band is offset from the non-ionizing UV continuum emission peaks. If this emission indeed corresponds to LyC radiation from the galaxy, it would be consistent with similar findings reported for LyC leakers at $z \gtrsim 3$ \citep{Ji20, Yuan24}, where the LyC signal is detected using high-resolution HST imaging. To the southeast of this emission, we note another comparably bright emission complex in the HST F150LP band. Although it lies within the stellar extent of the system, no counterpart is evident in the redder HST UV bands.

\subsection{Spectroscopic analysis and indicators} \label{sec:spec}
\begin{figure*}[ht!]
\nolinenumbers
\includegraphics[width=1\textwidth]{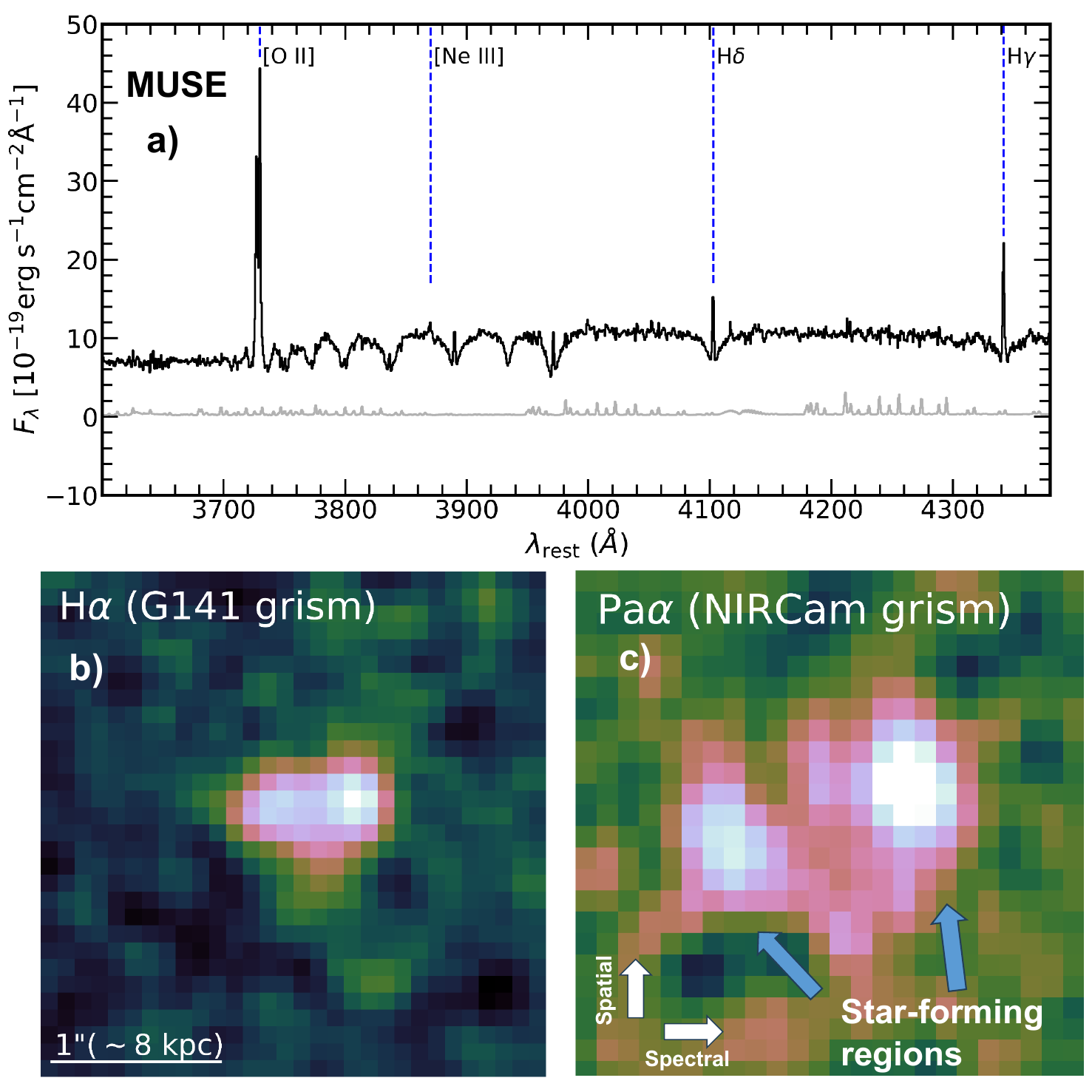}
\caption{Panel a) shows the MUSE 1D spectrum of MUSE ID 16 in the rest-frame, highlighting wavelengths redward of the [O II] doublet. The lines which we have used for analysis in this work are indicated with the blue dashed lines. The grey spectrum denotes the error spectrum. Panel b) presents the H$\alpha$ emission line map derived from HST G141 grism data. Panel c) displays the 2D JWST/NIRCam grism spectrum, zoomed in on the Pa$\alpha$ emission complex. White arrows indicate the spectral (increasing wavelength) and spatial directions.} 
\label{fig:spec}
\end{figure*}

The MUSE spectrum of MUSE ID 16 reveals a rich array of spectral features, including evidence for cool outflows, which we discuss in Section~\ref{sec:esc_theory}. In Panel a) of Figure \ref{fig:spec}, we present the portion of the spectrum redward of the [O II] doublet, as obtained from the MUSE HUDF survey \citep{Bacon23}. Using the calibration from \citet{Nagao06}, we estimate the metallicity based on the [Ne III]$\lambda$3869/[O II]$\lambda$3727 line ratio. Our results indicate a metallicity consistent with solar abundance for this object. Furthermore, using this ratio, we indirectly estimate the $[\text{O III}] \lambda5007 / [\text{O II}] \lambda3727$ (O32) ratio based on the empirical calibration from \citet{Witstok21}. We obtain an O32 value of $\sim$0.54, indicative of a low ionization state, typically not expected for LyC-leaking galaxies. However, recent studies have revealed a population of LyC leaker candidates exhibiting similarly low O32 values \citep{Bassett19,
Maulick24, Maulick25, Roy24}.
\par We also note prominent stellar Balmer absorption features in the MUSE spectrum that are characteristic of A-type stars \citep{GonzalezDelgado99, Pawlik18} which confirm the presence of an older stellar population within the object. We quantitatively measure the H$\delta$ and H$\gamma$ absorption equivalent widths using the emission-line masked spectrum, adopting the H$\delta_{\text{A}}$ and H$\gamma_{\text{A}}$ definitions from \citet{Worthey97}. By comparing these measurements to BPASS model predictions assuming a Salpeter IMF and solar metallicity \citep{Elridge17,Stanway18}, we estimate the age of the older stellar population to be approximately 0.8–1 Gyr. {Using the MUSE continuum and error spectra, we also measure the D4000 index \citep{Balogh99}. We adopt the standard definition, integrating the flux over 3850–3950 \text{\AA} for the blue continuum and 4000–4100 \text{\AA} for the red continuum. A Monte-Carlo propagation of the spectral uncertainties yields D4000 = $1.21 \pm 0.59$. Although the uncertainty is large, the measured value is still consistent with the presence of an underlying $\sim0.8-1$ Gyr stellar population inferred independently from the H$\delta_{\text{A}}$ and H$\gamma_{\text{A}}$ indices.}

\par We use the HST G141 grism spectrum to probe the H$\alpha$ emission from the galaxy. Panel b) of Figure \ref{fig:spec} shows the H$\alpha$ emission line map constructed from the G141 grism data. The reduction of the grism data was carried out using the grism redshift and line analysis software, \texttt{grizli}\footnote{\href{https://github.com/gbrammer/grizli/}{https://github.com/gbrammer/grizli/}} \citep{Brammer19}, following the methodology described in \citet{Maulick24,Maulick25}. For additional details, we refer the reader to \citet{Simons21,Simons23}. The peak of the H$\alpha$ emission is found to coincide approximately with the western star-forming complexes identified in Figure \ref{fig:stamp_images}. Using Gaussian fits to the 1D spectra generated with \texttt{grizli}, we estimate the H$\alpha$ line flux of the galaxy. An independent measurement from aperture photometry on the emission-line map yields a consistent flux value. Since the H$\alpha$ and the [N II]$\lambda\lambda$ 6548, 6583 lines are blended in the G141 grism spectrum, we apply a correction for the [N II] contamination ($\sim34\%$) using the empirical parameterization from \citet{Faisst18}, which depends on the stellar mass and redshift.

\par We also analyze JWST/NIRCam grism data in the F444W filter obtained from the FRESCO survey \citep{Oesch23}, which covers the Paschen-$\alpha$ (Pa$\alpha$) line in this galaxy. As a tracer of star formation, the Pa$\alpha$ line is less affected by dust, compared to H$\alpha$ or H$\beta$ \citep{Reddy23, Neufeld24}. The FRESCO data were reduced following the methodology outlined by \citet{Sun23}, using the associated publicly available workflow\footnote{\href{https://github.com/fengwusun/nircam\_grism}{https://github.com/fengwusun/nircam\_grism}}, which itself builds on the JWST science calibration pipeline \citep{Bushouse23}. We detect significant Pa$\alpha$ emission consistent with the spectroscopic redshift of the galaxy.

\par To isolate the pure emission-line signal, we apply an empirical median filtering technique using a boxcar kernel, following the procedure of \citet{Kashino23}, and subtract the continuum image obtained from the 2D spectrum. The resulting 2D grism spectrum, zoomed into the Pa$\alpha$ emission region, is shown in panel c) of Figure \ref{fig:spec}. Similar to the H$\alpha$ emission, the brightest pixel in Pa$\alpha$ is likely associated with the clumpy blue star-forming regions.

\par We extract the 1D spectrum from the reduced 2D NIRCam grism data, to estimate the Pa$\alpha$ line flux. Given the extended nature of the source, optimal extraction \citep{Horne86} was found to underestimate the line flux. We therefore adopt a simple box extraction method with an aperture width of 1$''$ in the spatial direction, accepting the associated trade-off of increased noise. The observed line fluxes for both H$\alpha$ and Pa$\alpha$ are reported in Table \ref{tab:MUSE16_prop}.

\subsection{Spectral Energy Distribution fitting}
\begin{figure*}[ht!]
\nolinenumbers
\includegraphics[width=1\textwidth]{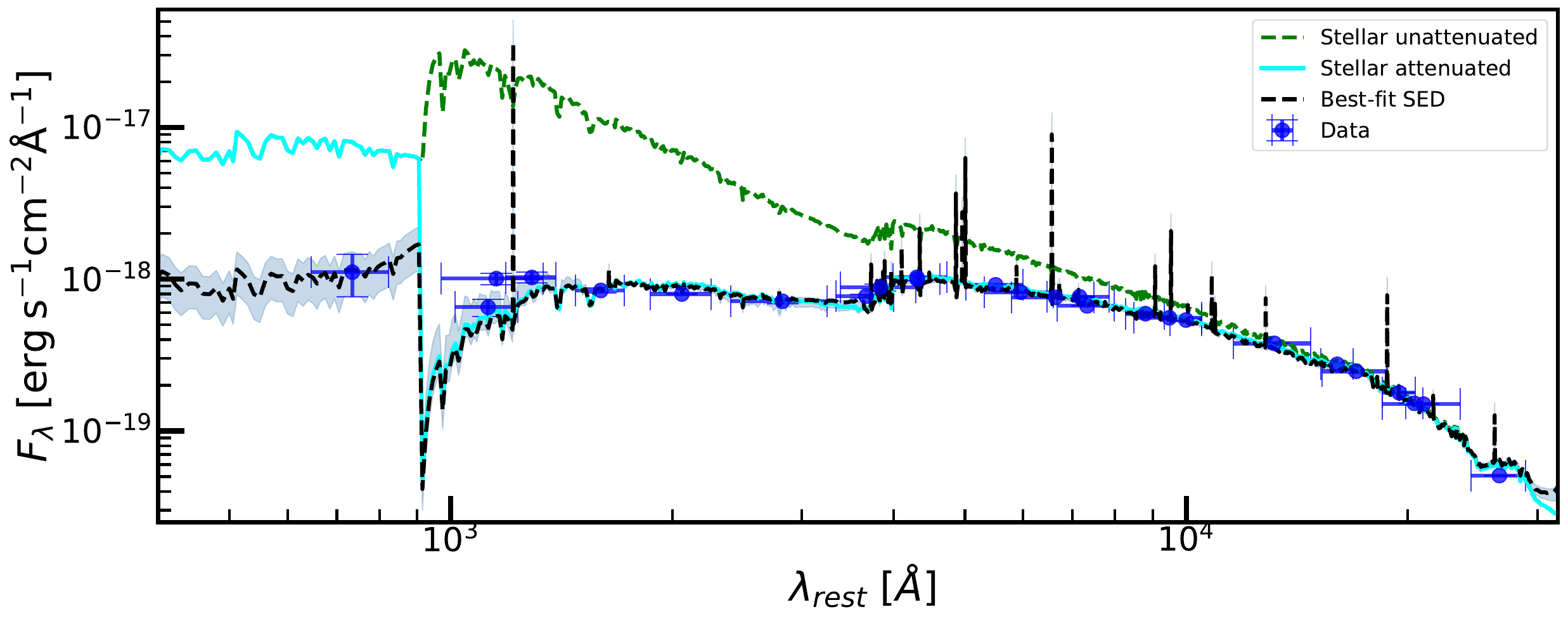}
\caption{The best-fit SED of 
MUSE ID 16, showing the observed model (black dashed), the intrinsic unattenuated stellar component (green dashed), and the dust-attenuated stellar component (cyan), overlaid with the measured photometric data points (blue circles). {The error in the x-axis denotes the effective width of the photometric broad bands}. {The blue shaded region represent the $16^{\text{th}}-84^{\text{th}}$ percentiles around the best-fit SED. The reduced $\chi^2$ of the best-fit is 0.72.}} 
\label{fig:SED}
\end{figure*}

We perform spectral energy distribution (SED) fitting of 
MUSE ID 16 using the code CIGALE \citep{Boquien19}. Prior to this, we carry out broadband photometry on the UVIT, HST (UVIS, ACS/WFC, and WFC3/IR), and JWST NIRCam imaging data. The filters used for this analysis are labeled in Figure \ref{fig:SED}. Photometric measurements are obtained using circular apertures of radius 1.2$''$, centered at the centroid coordinates of the NIRCam F444W image and have been corrected for Galactic extinction using the color excess from \citet{Schlafly11} and the \citet{Cardelli89} extinction curve. We additionally incorporate the flux measurement of 
MUSE ID 16 in the JWST/MIRI F560W band from the MIDIS catalog \citep{Ostlin25} into our fitting.
The UVIT fluxes are corrected for aperture losses using the PSF curve-of-growth derived from the AUDF South dataset \citep{Saha24}. Flux uncertainties are estimated from the associated weight maps.

We adopt the Bruzual and Charlot simple stellar population models (BC03, \citealt{Bruzual03}), a Salpeter initial mass function (IMF, \cite{Salpeter55}), and motivated by our spectroscopic analysis, implement the sfh2exp module to model the star formation history (SFH). This parameterization captures both an older underlying stellar population and a more recent starburst component. Based on our metallicity estimates from the Ne3O2 ratio, we fix both the stellar and gas metallicity to solar. The dust attenuation prescription and other module settings are adopted as in our previous analyses in \cite{Maulick25}. 
\par {To assess the sensitivity of the best-fit SED to the measured photometry, we also perform 1000 additional fits by perturbing the observed fluxes with Gaussian noise drawn from their respective flux uncertainties, and refitting each realisation using the same model grid.}
We present the best-fit spectral energy distribution (SED) in Figure \ref{fig:SED}, {with the 16th and 84th percentile SEDs from the 1000 noise-perturbed runs shown as the blue shaded region}, and summarize some of the derived physical properties in Table \ref{tab:MUSE16_prop}. \par {The best-fit model from the SED fitting yields D4000$=$1.15, consistent with the value derived independently from the MUSE spectrum.}
\par From the intrinsic stellar models corresponding to the best-fit SED, we estimate the ionizing photon production efficiency, defined as $\xi_{\rm{ion}}=\frac{\dot{N}^{int}_{\rm{LyC}}}{L_{1500}}$, where $\dot{N}^{int}_{\rm{LyC}}$ denotes the intrinsic ionizing photon production rate and $L_{1500}$ denotes the intrinsic luminosity density at rest-frame $1500\: \text{\AA}$ in units of $\text{erg}\  \text{s}^{-1} \text{Hz}^{-1}$ for the galaxy. {While the value of $\log_{10}(\xi_{\rm ion}/{\rm Hz\:erg}^{-1}) = 25.14$ for MUSE ID 16 lies close to the commonly adopted canonical value of 25.2 \citep{Robertson13}, it is slightly below the value of 25.29 obtained by extrapolating in redshift the $\log_{10}(\xi_{\rm ion}/{\rm Hz\:erg}^{-1})$–$z$ relation presented in \citet{Simmonds24} for star-forming galaxies (defined by $\log({\rm SFR}_{10\:{\rm Myr}}/{\rm SFR}_{100\:{\rm Myr}}) > -1$, a condition satisfied for MUSE ID 16 in our best fit SED model). The estimated value is, however, consistent with the $\log_{10}(\xi_{\rm ion}/{\rm Hz\:erg}^{-1})$–$M_{\rm UV}$ relation of \citet{Austin25}. We note that both relations exhibit substantial intrinsic scatter, and that they are primarily calibrated on higher-redshift ($z \gtrsim 3$) galaxy samples, with UV continuum slopes largely smaller than -2, and rely on SED-fitting methodologies that differ from the one employed here. }

\subsection{Dust attenuation} \label{sec:dust}
We estimate the nebular dust attenuation using the observed flux ratio of H$\alpha$ to Pa$\alpha$.  Assuming a gas temperature of $T = 10^4$K, we adopt an intrinsic H$\alpha$/Pa$\alpha$ ratio of 8.15 \citep{Ferland06}. To estimate the nebular reddening, ${E(B-V)_{\text{neb}}}$, we use the nebular attenuation curve from \citet{Reddy20}. This yields a nebular color excess of ${E(B-V)_{\text{neb}}} = 0.34 \pm 0.20$. Our best-fit SED, which adopts a modified Calzetti attenuation curve \citep{Calzetti00}, also reproduces a consistent nebular color excess of $\sim 0.30$. We note, however, that the flux measurements are derived from two different instruments, and potential systematic differences in their flux calibration remain unknown.

\par We subsequently calculate the star formation rate ($\text{SFR}_{\text{Pa}\alpha}$) for 
MUSE ID 16 using the dust-corrected Pa$\alpha$ luminosity and the calibration of \citet{Kennicutt98}, assuming a Salpeter IMF. 
\par The UV continuum slope ($\beta$) is known to correlate tightly with stellar dust content in galaxies \citep{Reddy18}. Using rest-frame UV photometry from three HST bands that approximately fall within the wavelength range defined by \citet{Calzetti94}, we fit a power-law of the form $F_{\lambda} \propto \lambda^{\beta}$ and derive a UV slope of $\beta = -0.31 \pm 0.07$, in good agreement with the value from our best-fit CIGALE SED model ($\beta \sim -0.34$). This is a relatively red UV continuum slope compared to typical LyC-emitting galaxies, which often exhibit bluer values \citep{Flury22a,Chisholm22}. It is worth noting, however, that the relatively red UV continuum slope we derive is based on integrated photometry. As discussed in Section \ref{sec:morph}, the star formation in this galaxy appears to be concentrated in clumpy regions. As a result, our measurement may be biased by contributions from dustier regions and older stellar populations across the galaxy. Studies of lensed galaxies, such as \citet{Bolamperti23}, have shown that UV continuum slopes measured in individual star-forming clumps tend to be significantly bluer than those derived from integrated light. {To investigate this effect directly, we apply the Python implementation of Source Extractor \citep{Bertin96}, \textit{sep} \citep{Barbary16}, to the F336W image, where the emission is dominated by the UV-bright clumps. Using appropriate deblending parameters, we are able to isolate the northwestern and southwestern clumps as separate detections (marked by the cyan arrows in Figure~\ref{fig:stamp_images}). We then measured the UV continuum slopes of these individual clumps using the same set of filters adopted for the integrated galaxy, but employing circular apertures of radius $0.12''$, that is approximately the FWHM of the relevant HST images, and applying aperture corrections. This choice minimizes contamination from surrounding diffuse emission and neighboring clumps. The resulting UV slopes are bluer than that of the integrated galaxy, with the northwestern clump having $\beta_{\text{obs}} = -0.87 \pm 0.18$ and the southwestern clump $\beta_{\text{obs}} = -1.03 \pm 0.03$.}

Using the empirical relation between the UV continuum slope ($\beta$) and the stellar continuum color excess ${E(B-V)_{\text{stellar}}}$ derived for a Calzetti attenuation curve \citep{Reddy18}, we estimate ${E(B-V)_{\text{stellar}}} = 0.49 \pm 0.01$, for the whole galaxy. Interestingly, this value is larger than the nebular color excess, which is contrary to the commonly held view that star-forming regions typically experience more dust attenuation than the stellar continuum \citep{Calzetti01}.

\subsection{Escape fraction}
We estimate the escape fraction of Lyman-continuum (LyC) photons from the galaxy using the observed F154W flux, the dust-corrected H$\alpha$ luminosity, and our best-fit SED model. The escape fraction definition we adopt here is closest to the absolute escape fraction, expressed as:

\begin{equation}
f_{\rm{esc,LyC}} = \frac{\dot{N}^{\rm{esc}}_{\rm{LyC}}}{\dot{N}^{\rm{esc}}_{\rm{LyC}} + \dot{N}^{\rm{non-esc}}_{\rm{LyC}}},
\label{eq:fesc}
\end{equation}

where $\dot{N}^{\rm{esc}}_{\rm{LyC}}$ [s$^{-1}$] is the rate of ionizing photons escaping the galaxy (corrected for IGM absorption), and $\dot{N}^{\rm{non-esc}}_{\rm{LyC}}$ [s$^{-1}$] is the rate of ionizing photons absorbed internally, either through recombination or dust absorption.

We estimate $\dot{N}^{\rm{esc}}_{\rm{LyC}}$ from the observed F154W flux following the methodology of \citet{Maulick25}. Attenuation by the IGM is implicitly incorporated using the IGM transmission models in \citet{Dhiwar24}, from which we adopt the mean IGM transmission at $z \sim 1.2$ from the \citet{Inoue14} set of column density distribution functions. It is important to note that this use of a mean transmission results in a likely overestimation of the numerator in Equation \ref{eq:fesc}, as the detection of LyC photons likely favors sightlines with higher-than-average IGM transparency \citep{Bassett21}. On the other hand, since the F154W filter probes only part of the ionizing spectrum, this estimate is an approximation of the true $\dot{N}^{\rm{esc}}_{\rm{LyC}}$.

To calculate the non-escaping ionizing photon rate $\dot{N}^{\rm{non-esc}}_{\rm{LyC}}$, we use the dust-corrected H$\alpha$ luminosity under the assumptions of Case B recombination, a temperature of $T = 10^4$ K, and an electron density of $n_e = 100$ cm$^{-3}$. Following \citet{Ferland06}, this rate is given by:

\begin{equation}
 \dot{N}^{non-esc}_{\rm{LyC}} [\:\text{s}^{-1}]=7.28 \times 10^{11} L_{\rm{int},\rm{H}\alpha} [\text{erg}\: \text{s}^{-1}],
\label{eq:N_nonesc}
\end{equation}

where $L_{\rm{int, H\alpha}}$ is the intrinsic (dust-corrected) H$\alpha$ luminosity. The escape fraction derived from Equations~\ref{eq:fesc} and \ref{eq:N_nonesc} is listed as $f^{\rm{H\alpha}}_{\rm{esc}}$ in Table~\ref{tab:MUSE16_prop}.

\par
As an alternative approach, we also estimate $f_{\text{esc}}$ from our best-fit SED, where the escape fraction is treated as a free parameter. This estimate corresponds to the escape of ionizing photons across the entire ionizing spectrum. However, we caution that in CIGALE’s modeling framework, ionizing radiation is assumed to be dust-transparent. We report this SED-based value as $f_{\rm{esc,SED}}$ in Table~\ref{tab:MUSE16_prop}.

\section{Discussion} \label{sec:discussion}
\subsection{The escape of ionizing photons from a stratified disk} \label{sec:esc_theory}

Analytical work by \citet{Dove94} for stratified gas density profiles (e.g. a Gaussian profile) demonstrated that HII regions embedded in galactic disks are more likely to be density-bounded in the direction perpendicular to the disk plane (see their Figure 1). Superbubbles generated by supernova-driven mechanical feedback in star-forming regions tend to expand preferentially in the vertical direction \citep{MacLow88}. These expanding bubbles carve dynamic chimneys or low-density channels through which the LyC photons can escape, a scenario supported by subsequent modeling in \citet{Dove00}. However, the shells of these evolving superbubbles also tend to trap the ionizing radiation, thus decreasing the escape fraction. The preferential escape of LyC photons along the galactic poles has also been confirmed in high-resolution simulations such as those of \cite{Gnedin08}.

At face value, the integrated properties of 
MUSE ID 16 do not appear particularly conducive to LyC escape. For instance, the galaxy exhibits a relatively low ionization state, as indicated by its Ne3O2 ratio, and a red UV continuum slope.
From a morphological perspective, one of the key diagnostics of strong feedback often linked to LyC escape is the galaxy's compactness, often quantified by its half-light radius, $r_{50}$, and its star-formation rate surface density, $\Sigma_{\text{SFR}}$ $[\text{M}_{\odot}\:\text{yr}^{-1}\:\text{kpc}^{-2}]$ \citep{Cen20,Carr25}. 

\begin{figure*}[ht!]
\nolinenumbers
\includegraphics[width=1\textwidth]{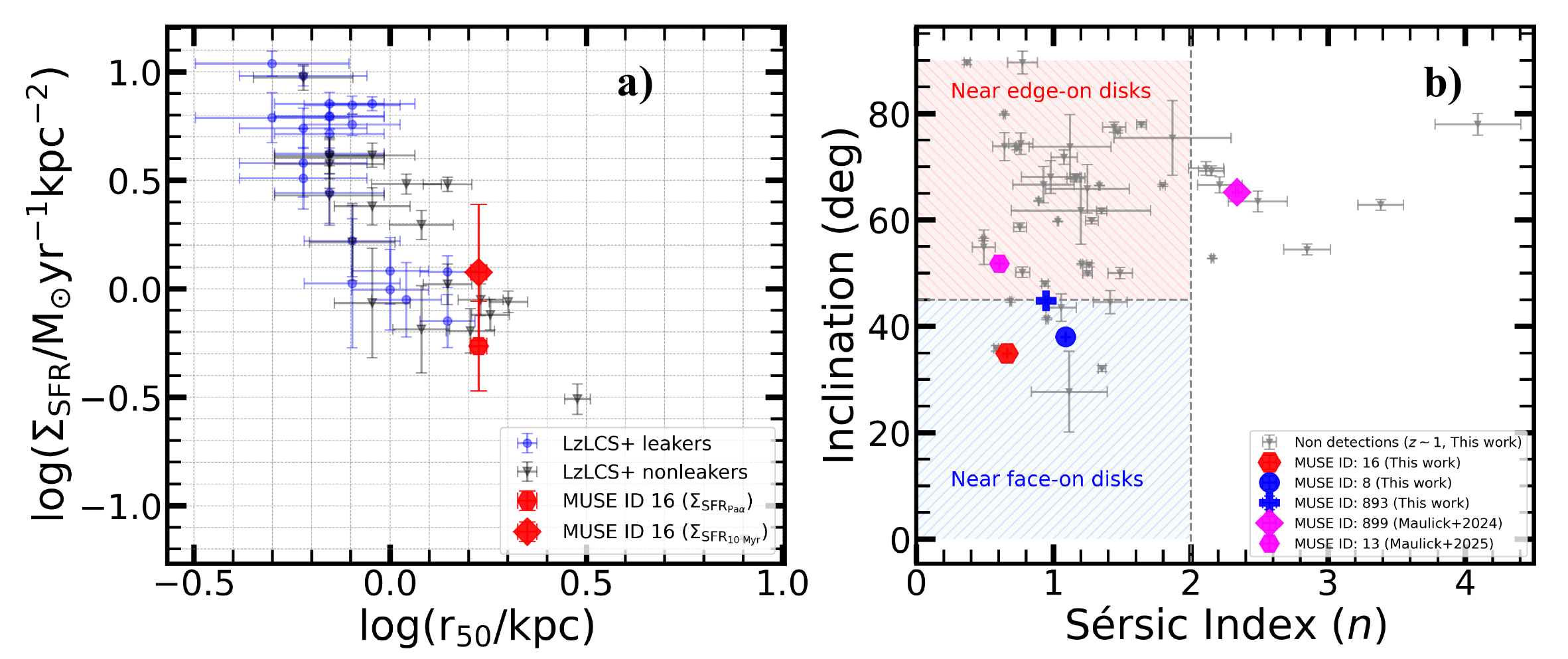}
\caption{Panel a) displays the star-formation rate surface density, ($\log(\Sigma_{\rm{SFR}}/\rm{M}_{\odot}\:\rm{yr}^{-1}\:\rm{kpc}^{-2})$, versus the UV half-light radius, ($\log(r_{50}/\rm{kpc})$), for 
MUSE ID 16 (red markers), compared to the low-redshift Lyman-continuum leaker pulus sample (LzLCS+) from \citet{Flury22a} (blue and grey markers), with updated structural measurements adopted from \citet{LeReste25}. Panel b) shows the distribution of LyC detections (blue, red, and magenta markers) and non-detections (inverted grey triangles) from our sample (Section~\ref{sec:selection}) in the inclination–Sérsic index plane. The axis ratios and Sérsic indices are taken from \citet{vanderWel12}. Galaxies are classified as near face-on or near edge-on disks according to the criteria described in Section~\ref{sec:inclination}.} 
\label{fig:sigmasfr_r50}
\end{figure*}

In panel a) of Figure~\ref{fig:sigmasfr_r50}, we place 
MUSE ID 16 on the $\Sigma_{\text{SFR}}$–$r_{50}$ plane, comparing its properties with the low-redshift Lyman-continuum leaker plus sample (LzLCS+), which includes galaxies from \citet{Flury22a} and references therein. We estimate the UV half-light radius of 
MUSE ID 16 using the HST F336W band, while for the LzLCS+ sample, we adopt updated HST-based measurements from \citet{LeReste25} for $\Sigma_{\text{SFR}}$ and $r_{50}$.

To calculate the $\Sigma_{\text{SFR}}$ of 
MUSE ID 16, we use two independent tracers of recent star formation: the dust-corrected Pa$\alpha$ luminosity and the SED-inferred SFR averaged over the past 10 Myr, both of which are sensitive to the young stellar populations responsible for LyC production.

\begin{figure*}[ht!]
\nolinenumbers
\includegraphics[width=0.88\textwidth]{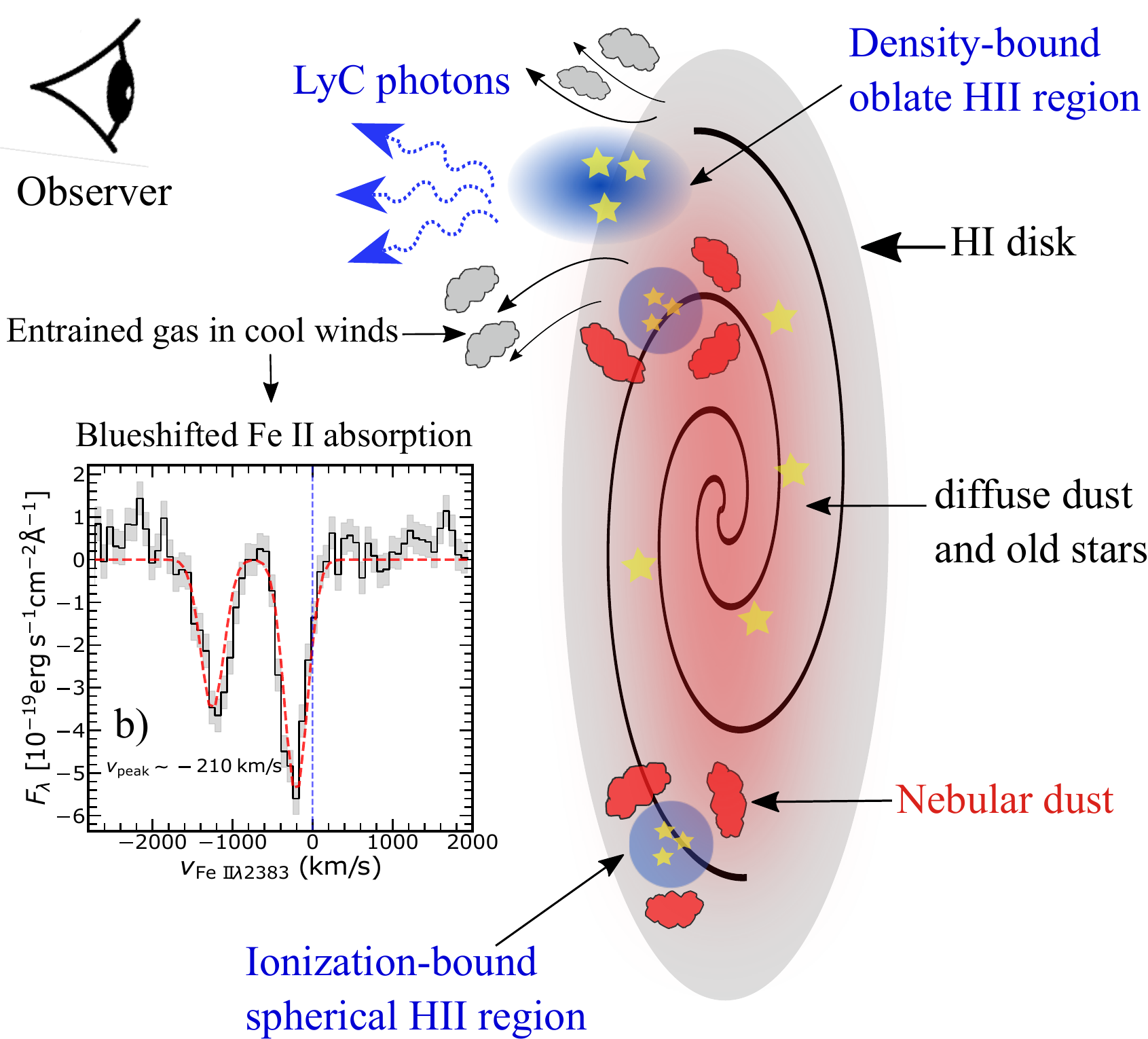}
\caption{Simplified cartoon schematic illustrating the escape of ionizing photons from a star-forming spiral disk galaxy viewed nearly face-on, representing the plausible scenario for 
MUSE ID 16. LyC photons are depicted emerging in a conical fashion from a density-bounded, oblate HII region. In contrast, ionization-bounded star-forming regions remain embedded within dust and neutral gas. Panel b) highlights the blueshifted Fe II$\lambda$2383 absorption relative to the systemic redshift determined from the [O II] doublet, indicative of cool galactic winds and outflowing, entrained gas along the line of sight. Note that the CGM and molecular clouds are not included in this schematic.} 
\label{fig:schematic}
\end{figure*}

The LzLCS+ sample reveals that galaxies with compact sizes ($r_{50}<1$ kpc) and high star-formation rate surface densities ($\Sigma_{\text{SFR}}>1\:\text{M}_{\odot}\:\text{yr}^{-1}\:\text{kpc}^{-2}$) are dominated by confirmed LyC leakers. In contrast, the region of lower $\Sigma_{\text{SFR}}$ and more extended sizes where 
MUSE ID 16 resides hosts a mix of both leakers and non-leakers. Nonetheless, 
MUSE ID 16 falls within the general range of $\Sigma_{\text{SFR}}\sim0.1-1\:\text{M}_{\odot}\:\text{yr}^{-1}\:\text{kpc}^{-2}$, a regime typically associated with the minimum star-formation rate surface density required for the launching of galactic winds and outflows \citep{Heckman02,Newman12}. Supporting this, the MUSE spectrum of 
MUSE ID 16 shows signs of such feedback. We detect blueshifted Fe II$\lambda$2383 absorption, indicative of cool, outflowing entrained gas \citep{Heckman02,Erb12}. The peak of this absorption feature is offset by nearly 200 km/s from the systemic redshift (Panel b, Figure~\ref{fig:schematic}), consistent with wind velocities observed in other massive star-forming galaxies with similarly red UV continuum slopes \citep{Erb12}.

The orientation of 
MUSE ID 16 could then play a significant role in its observed LyC leakage. To explore this, we first make a rough estimate of its effective Strömgren radius under standard assumptions. Assuming a gas temperature of 10,000 K and an electron density of $n_{\rm{e}} \sim 100\:\rm{cm}^{-3}$ which is consistent with that inferred from our [O II] doublet ratio measured in the MUSE spectrum and using PyNeb \citep{Luridiana15}. We use the classic Strömgren sphere formalism \citep{Dove00,Ferland06}, assuming spherical geometry:

\begin{equation}
R_{\text{S}} \approx \left(\frac{3Q(\text{H}^0)}{4\pi n_{\text{H}}^2 \alpha_{B}}\right)^{1/3} = 152\:\text{pc} \: Q_{49}^{1/3} \left(\frac{0.3\:\text{cm}^{-3}}{n_{\text{H}}}\right)^{2/3},
\label{eq:Stromgren}
\end{equation}

where $Q(\text{H}^0)$ is the hydrogen-ionizing photon production rate in s$^{-1}$, $Q_{49}=10^{49} \:\text{s}^{-1}$, $n_{\text{H}}$ is the hydrogen number density (assumed to be approximately equal to $n_{\text{e}}\:[\text{cm}^{-3}]$), and $\alpha_{\rm{B}} \:[\text{cm}^{3}\:\text{s}^{-1}]$ is the Case B recombination coefficient. Using Equation~\ref{eq:N_nonesc} to estimate $Q(\text{H}^0)$ from the dust-corrected H$\alpha$ luminosity, we obtain an effective Strömgren radius of approximately 131 pc.

We refer to this as the effective Strömgren radius because it relies on the simplifying assumption that the entire H$\alpha$ emission arises from a single, compact star-forming region. However, the morphology seen in Figure~\ref{fig:stamp_images}, and especially panels (b) and (c) of Figure~\ref{fig:spec} clearly indicates that the star-forming regions are resolved and thus spatially extended.

The relation between H$\alpha$ luminosity and size can offer insight into whether these regions are density-bounded. Although we defer a detailed analysis of this for future work, it is worth noting that studies \citep[e.g.,][]{Wisnioski12,Cosens18} have found that high-redshift H$\alpha$-emitting star-forming regions with $\Sigma_{\text{SFR}} > 1\:\text{M}_{\odot}\:\text{yr}^{-1}\:\text{kpc}^{-2}$ tend to deviate from the classical $L_{\text{H}\alpha} \propto r_{\text{H}\alpha}^{3}$ scaling expected from Strömgren theory. This deviation may reflect elevated star-formation activity at high redshifts, leading to regions that are density-bounded, especially in galaxies with a disk-like geometry \citep{Wisnioski12}.

As discussed earlier, in a stratified disk geometry, ionizing photons are expected to escape preferentially in the direction perpendicular to the disk plane. In the idealized case of a dust-free, pure hydrogen nebula where absorption occurs only via photoelectric processes, the vertical distribution of neutral hydrogen can be modeled by a Gaussian profile: \begin{equation} \label{eq:HI_gauss}
    n_\text{H}(Z)=n_0\:\text{exp}\big[-Z^2/2h^2\big]
\end{equation}, where $Z$ is the height above the midplane, $h$ is the HI scale height and $n_0$ is the midplane density in $\text{cm}^{-3}$. In such a setup, LyC photons are predicted to escape through a cone-like structure originating at the source of the ionizing radiation (e.g., an O or B star). The opening angle of this ionization cone increases with the ionizing photon luminosity of the source, meaning that stronger ionizing sources carve out wider escape cones \citep{Dove94, Benson13}. Outside this cone, the escape fraction drops rapidly to zero \citep{Benson13}. This framework assumes that the ionizing photons are emitted from stars that lie in the plane of the disk. However, a non-negligible fraction of O and B stars, known as runaway stars, can attain peculiar velocities exceeding 30 km/s as observed in the Milky Way \citep{Blaauw61}, allowing them to travel distances of $\gtrsim150$ pc within 5 Myr, effectively displacing them from their dense birth clouds. By residing in lower-density environments, these stars can emit LyC photons along clearer sightlines, thereby potentially boosting the overall escape fraction \citep{Conroy12, Kimm14}. It is worth noting that \citet{Dove94} originally proposed that LyC escape in such a configuration within the local Universe is only possible from diffuse HII regions with a hydrogen number density of $n_{\text{H}}\lesssim1\:\text{cm}^{-3}$, which is significantly larger than our assumed $n_{\text{H}}\sim100\:\text{cm}^{-3}$. However, the impact of this difference on the Strömgren geometry may be mitigated by the increase in the ionizing photon production rate, $Q(\text{H}^0)$, as we move closer to the peak of the cosmic star-formation history.

The {HI vertical ($Z$ in Equation \ref{eq:HI_gauss}) distribution} in 
MUSE ID 16 remains unconstrained. More broadly, the assumption of a stratified HI disk at $z\sim1$ is an extrapolation from the local Universe and has not been observationally verified. While results from ionized gas kinematics suggest that a significant fraction of star-forming galaxies are mature rotating disks at this redshift \citep{Wisnioski15,DiTeodoro16}, the vertical structure and stratification of the neutral atomic gas remain unknown. For context, the HI scale height in the Milky Way, unlike the idealized, infinite slab approximation described by Equation~\ref{eq:HI_gauss} is observed to vary with galactocentric radius. In particular, disk flaring causes the scale height to increase with distance from the Galactic center \citep{Kalberla08,Saha09}. At the solar neighborhood, the HI scale height near the midplane is approximately 150 pc \citep{Kalberla08}, which is comparable to the effective Strömgren radius of 131 pc estimated earlier. The Milky Way scale height remains below 1 kpc out to galactocentric distances of about 20–25 kpc \citep{Kalberla08,Saha09}.

We have thus far largely neglected the role of dust. In simplified models where dust and gas are perfectly mixed, the dust absorption cross-section is typically assumed to be about four orders of magnitude lower than the photoionization cross-section in low-metallicity systems ($Z<0.3\:Z_{\odot}$), based on extrapolations from the SMC extinction curve \citep{Gnedin08}. However, given that 
MUSE ID 16 is expected to have a near-solar metallicity (Section \ref{sec:spec}), dust extinction could become comparable to photoelectric absorption \citep{Kakiichi21}. Conversely, it is also plausible that the low-density channels through which ionizing photons escape are cleared of both gas and dust \citep{Reddy16}, or that LyC photons originate from a population of completely unobscured sources \citep{Gnedin08}.
\par Bringing together the key points discussed above, we illustrate a simplified representation in the schematic cartoon shown in Figure~\ref{fig:schematic}. This depicts a nearly face-on view of the spiral galaxy, highlighting the escape of LyC photons from a density-bounded, oblate HII region consistent with the framework proposed by \citet{Dove94}, while also showing the presence of ionization-bounded HII regions. This picture is motivated by Figure~\ref{fig:stamp_images}, which shows two UV-bright star-forming clumps (marked with cyan arrows in Figure \ref{fig:stamp_images}), as well as by the detection of CO emission from 
MUSE ID 16 \citep{Boogard19}, with CO contours concentrated near the northern clump. This suggests the likely presence of significant neutral gas along the line of sight to the northern star-forming region. Using the CO size and molecular gas mass estimate from \citet{Aravena19}, we derive a molecular hydrogen column density of approximately $N_{\text{H}_2}=(3/4)(\Sigma_{\text{mol}}/m_{\text{H}})\sim 10^{22}\:\text{cm}^{-2}$ \citep{Kakiichi21}, suggesting that any neutral HI relevant to LyC escape along this line of sight could also be sufficiently dense to cause significant attenuation. We therefore speculate that the LyC photons are more likely escaping from the southern clump or from a region offset from both star-forming sites, {as suggested by the spatial offset between the eastern F150LP emission peak and the peaks of the non-ionizing UV continuum (Figure~\ref{fig:stamp_images})}. The mechanical feedback from the northern clump may yet play a role in shaping the interstellar medium, potentially contributing to the formation of channels through which LyC photons escape similar to the scenario proposed for J1044+0353, where a $\gtrsim 10$ Myr stellar population is thought to provide the necessary feedback for LyC leakage from the starburst region \citep{Herenz25}. 
\par While the exact origin of LyC emission may differ for the other face-on LyC leakers (possibly AGN) shown in Figure~\ref{fig:MUSEID8_893}, the general arguments involving stratified geometry and viewing angle are likely applicable to them as well.

\subsection{Viewing angle and LyC detection} \label{sec:inclination}
In the previous subsection, we presented primarily theoretical arguments suggesting that in disk galaxies, LyC photons are more likely to escape perpendicular to the plane of the disk.

Here, we investigate whether LyC detections in our whole sample (defined in Section~\ref{sec:selection}) are more prevalent in face-on disk galaxies compared to edge-on systems. For this, we include both detections and non-detections from Section~\ref{sec:selection}, identifying disk galaxies using Sérsic indices from \citet{vanderWel12}, limited to those with reliable structural fits (i.e., flag = 0 in their catalog). This brings the non-detection sample of 92 (Flag 1 in Section~\ref{sec:selection}) down to 70. To ensure we are selecting likely star-forming systems with reliable redshifts among these 70 non-detections, we further restrict this sample to galaxies exhibiting significant HI recombination emission. Specifically, we require S/N > 3 in either H$\alpha$ (cross-matched from the CLEAR survey \citealt{Simons23}) or H$\gamma$ (from the MUSE HUDF DR2 catalog \citealt{Bacon23}), reducing the sample of non-detections from 70 to 50 galaxies.
We classify the non-detections and the LyC leaker candidates with Sérsic index $n<2$ as disks. We convert the observed axis ratios into inclination angles using the classical Hubble formula \citep{Hubble26}: $\text{cos}^2(i)=(q^2-q_0^2)/(1-q_0^2)$, where where $q=b/a$ is the observed axis ratio and for simplicity we adopt $q_0=0.2$ as the intrinsic thickness of an edge-on disk \citep{Padilla08}. In Panel b) of Figure~\ref{fig:sigmasfr_r50}, we highlight the subset of disk galaxies ($n<2$) and divide them into two inclination regimes: 'near face-on' for those with $i<45^{\circ}$, and 'near edge-on' for those with $i>45^{\circ}$.
\par Based on our classification, we identify 44 disk galaxies ($n<2$) in the sample, of which 10 fall into the ‘near face-on’ category and 34 into the ‘near edge-on’ category. Among the face-on disks, 3 out of 10 ($30\%$) are detected in LyC, as presented in this work; this fraction becomes 1 out of 8 ($\sim12\%$) if we exclude MUSE IDs 8 and 893 due to their AGN nature. In contrast, only 1 out of the 34 ($\sim3\%$) edge-on disks shows LyC detection (object reported in \cite{altMaulick25}). We highlight these findings in Panel b of Figure \ref{fig:sigmasfr_r50}.
At face value, this result appears to support the hypothesis that LyC photons are more likely to escape along directions perpendicular to the disk plane. However, several caveats must be considered before drawing strong conclusions, particularly in the context of anisotropic LyC escape from disks in the EoR. First, our sample of face-on galaxies is small and thus susceptible to low-number statistics. While we have limited our comparison to galaxies that are likely star-forming based on the detection of HI recombination emission lines, this represents only a first-order selection. With a larger sample, a more robust analysis should involve comparing galaxies within matched bins of additional properties, such as stellar mass, specific star formation rate, or non-ionizing UV luminosity, to better control for underlying population differences.

While light profiles may appear disk-like and recent studies suggest that roughly 45\% of galaxies in the EoR show disk-like morphologies \citep{Sun24}, this alone does not determine LyC escape. Ultimately, the escape of ionizing photons is governed by the spatial distribution of young stars, neutral gas, and their associated feedback processes, along with the possible influence of AGN, rather than by stellar morphology inferred from light profiles alone, which remains uncertain for galaxies in the EoR. Furthermore, most LyC leakers at low redshift ($z\sim0.3$–0.4), such as those in the LzLCS+ sample, have been found to be morphologically irregular \citep{LeReste25}. Notably, MUSE IDs 13 and 899 in our sample are also irregular systems (based on visual inspection), although the former is formally classified as a ‘near edge-on disk’ by our Sérsic index and inclination-based criteria. This highlights that our classification scheme is purely morphological and does not capture the underlying dynamical or gas structural complexity. As such, it may only hint at trends in LyC detectability from disk-like systems.
\par It is also interesting to note that recently \cite{Lorenz23} show that the dust attenuation traced by both recombination lines and the UV continuum slope for a sample of galaxies between redshifts 1.3 and 2.6, shows no dependence on the viewing angle. They favor a model (see their Figure 9) in which the nebular dust of star-forming regions dominates the dust attenuation of star-forming galaxies rather than the diffuse dust in the ISM. Our result may then offer an interesting contrast: If dust is indeed a major sink for LyC photons and if these photons preferentially escape perpendicular to the disk plane from density-bounded regions, then it is plausible that such regions are cleared of dust which in turn enables relatively unobscured LyC escape along those sightlines. This may explain the inclination dependence we observe, highlighting the role of geometry and feedback in shaping LyC visibility.

\section{Summary} \label{sec:summary}
In this work, we examined a sample of 286 galaxies at $z \sim 1$ from the archival MUSE HUDF survey, in combination with data from the AUDFs survey, and identified three new LyC leaker candidates, all of which are nearly face-on spiral systems. One of these systems (MUSE ID 16) shows no evidence of AGN activity and is characterized by a red UV continuum slope, ($\beta_\text{obs}\sim-0.3$), a high stellar mass ($\text{log}_{10}(\text{M}_{*}/\text{M}_{\odot})\sim10.33$), a low ionization state, and an underlying old ($\sim1$ Gyr) stellar population. We explore the conditions that enable LyC escape from this galaxy and propose a scenario in which a compact, localized star-forming region, either intrinsically unobscured or situated within a low-density channel sculpted by mechanical feedback from itself and/or nearby ($\lesssim1$ kpc) star-forming regions facilitates the escape of ionizing photons along this favorable sightline. Such a configuration may explain the high inferred escape fraction ($\gtrsim0.3$), which exceeds predictions from earlier theoretical models based on homogeneous ISM assumptions. Finally, we find tentative evidence that galaxy orientation affects LyC detectability, with face-on systems more likely to exhibit leakage. This suggests that anisotropic escape, driven by geometry and feedback, may be especially important in extended, low-ionization systems.

Some of the data presented in this article were obtained from the Mikulski Archive for Space Telescopes (MAST) at the Space Telescope Science Institute. The specific observations analyzed can be accessed via the following \dataset[doi:10.17909/w12m-m046]{https://archive.stsci.edu/doi/resolve/resolve.html?doi=10.17909/w12m-m046}.

\textit{Acknowledgements:} We thank Edmund Christian Herenz for useful discussions. We gratefully acknowledge the MUSE HUDF, JADES, and FRESCO teams for publicly releasing their data, without which this work would not have been possible.

\facilities{AstroSat (UVIT), HST (WFC3, ACS-WFC, WFC3-UVIS), JWST (NIRCam, MIRI), VLT:Yepun (MUSE)}


\software{astropy \citep{2013A&A...558A..33A,2018AJ....156..123A},  
          CIGALE \citep{Boquien19}, photutils \citep{Bradley23}, \texttt{grizli} \citep{Brammer19}, \textit{sep} \citep{Barbary16}
          }


\appendix
\section{Face-on AGN LyC leaker spiral galaxies} \label{sec:AGN_leakers}
\begin{figure*}[ht!]
\nolinenumbers
\includegraphics[width=1\textwidth]{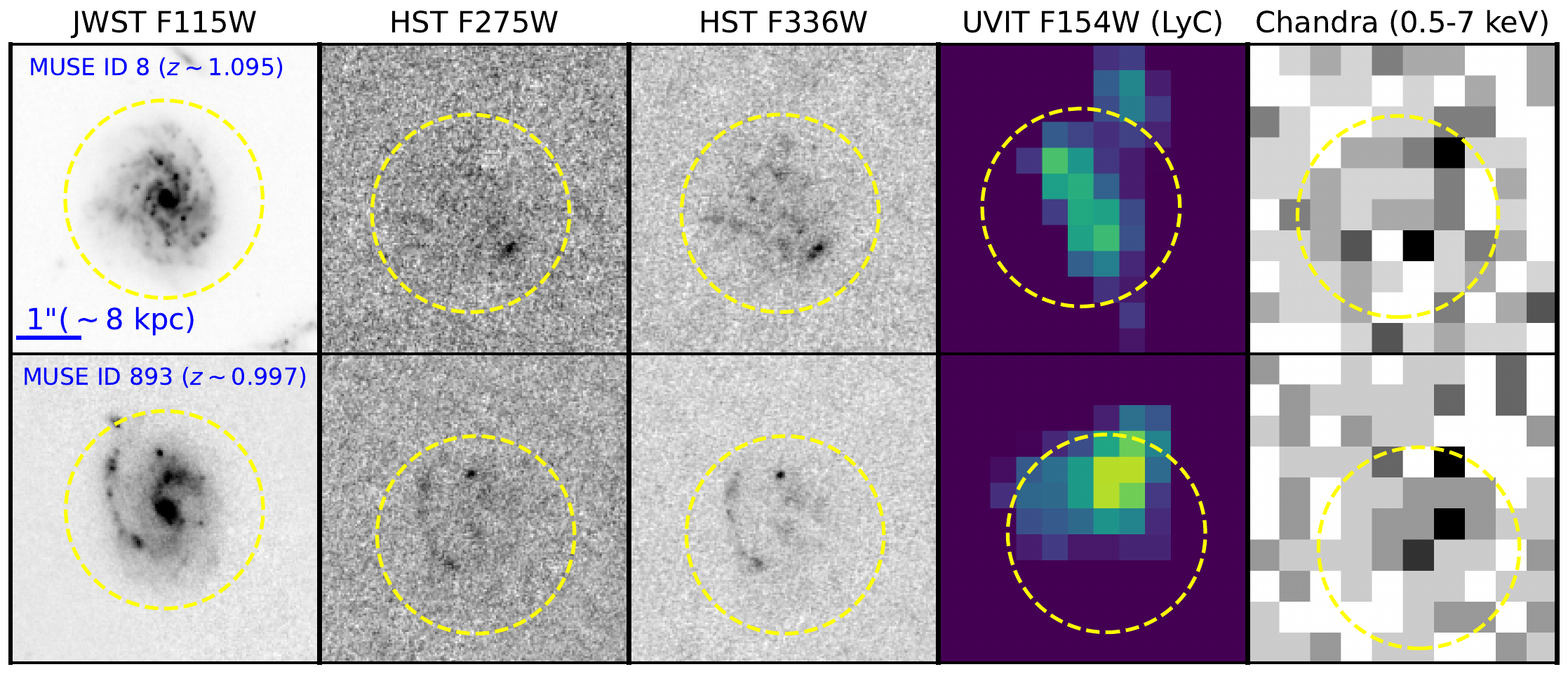}
\caption{Multiband imaging of the AGN LyC leaker candidates, MUSE IDs 8 and 893, at redshifts 1.095 and 0.997 respectively. From left to right, the panels show: JWST NIRCam F115W, HST WFC3/UVIS F275W, HST WFC3/UVIS F336W, UVIT F154W (probing the LyC), and Chandra X-ray imaging covering the full 0.5–7 keV energy range. The yellow dashed circle marks an aperture of radius 1.2$''$ centered on the NIRCam band centroid of the objects. The F154W images have been smoothed with a Gaussian kernel with an FWHM of 3 pixels.} 
\label{fig:MUSEID8_893}
\end{figure*}

As discussed in Section~\ref{sec:selection}, MUSE IDs 8 and 893, at redshifts 1.095 and 0.997 respectively, likely host AGN activity, as indicated by their X-ray detections and broad H$\alpha$ emission features. These two galaxies are displayed in Figure~\ref{fig:MUSEID8_893}. Their spiral-like morphologies are clearly visible in the JWST NIRCam F115W images.

Similar to 
MUSE ID 16 (Figure~\ref{fig:stamp_images}), both galaxies exhibit clumpy star-forming regions in the bluest HST bands, offset from the central nucleus. In particular, MUSE ID 893 displays an intriguing alignment: the observed LyC emission (third panel from the left in Figure~\ref{fig:MUSEID8_893}) appears coincident with the brightest of these star-forming clumps, as seen in the HST F275W and F336W bands. This spatial correspondence could suggest that this clump is the origin of the LyC photons. However, given the coarse spatial resolution of UVIT, this interpretation remains uncertain.

In contrast to 
MUSE ID 16, neither MUSE ID 8 nor MUSE ID 893 shows evidence of outflowing material in their rest-UV spectra, as inferred from the lack of Fe II$\lambda$2382 blueshifted absorption.

\bibliography{face_on_lyc_arxiv}
\bibliographystyle{aasjournal}



\end{document}